\documentclass[nopubldata]{lpor2014}
\usepackage{graphicx}  
\usepackage{dcolumn}   
\usepackage{bm}        
\usepackage{amssymb}   
\usepackage{amsmath}
\usepackage{amsfonts}
\usepackage{gensymb}
\usepackage{siunitx}
\usepackage{braket} 
\usepackage{caption}

\graphicspath{{figs/}}
\begin{document}
%
\titlefigure{Fig0}

\abstract{Open-access microcavities are emerging as a new approach to confine and engineer light at mode volumes down to the $\lambda^3$ regime. They offer direct access to a highly confined electromagnetic field while maintaining tunability of the system and flexibility for coupling to a range of matter systems. This article presents a study of coupled cavities, for which the substrates are produced using Focused Ion Beam milling. Based on experimental and theoretical investigation the engineering of the coupling between two microcavities with radius of curvature of \SI{6}{\micro\meter} is demonstrated. Details are provided by studying the evolution of spectral, spatial and polarisation properties through the transition from isolated to coupled cavities. Normal mode splittings up to 20 meV are observed for total mode volumes around $10 \times \lambda^3$. This work is of importance for future development of lab-on-a-chip sensors and photonic open-access devices ranging from polariton systems to quantum simulators.
}

\title{Spectral engineering of coupled open-access microcavities}
%
\titlerunning{}
\author{L.C.~Flatten\inst{*}, A.A.P.~Trichet\inst{}, J.M.~Smith\inst{}}
%
\authorrunning{L.C.~Flatten et al.}
%
\institute{%
Department of Materials, University of Oxford, Parks Road, Oxford OX1 3PH, United Kingdom
}
%
\mail{\email{lucas.flatten@materials.ox.ac.uk}}
%
%
\maketitle

\section{Introduction}
\label{sec::intro}
Optical microcavities, owing to their small confinement volume, serve to enhance the light-matter interaction. They significantly improve the control and sensing capacities for electronic systems situated in the confined region. Epitaxial and monolithic microcavities have made substantial contributions to various fields such as quantum optics \cite{vahala_optical_2003}, lasing \cite{ward_wgm_2011}, optomechanics \cite{kleckner_sub-kelvin_2006} and optical sensing \cite{vollmer_review_2012}. Building on improvements in the fabrication methods of these devices, it has been possible to move beyond the single cavity design towards coupled cavities \cite{bayer_optical_1998, vasconcellos_spatial_2011} and on-chip photonic circuits \cite{sato_strong_2012,boriskina_photonic_2010,obrien_photonic_2009}. As striking examples, coupled cavities have been used to realise highly efficient sources of entangled photon pairs based on semiconductor quantum dots \cite{dousse_ultrabright_2010}, macroscopic Josephson oscillations of exciton polaritons \cite{JosephsonOscillations_BLoch_2013} and photonic graphene \cite{jacqmin_direct_2014,nalitov_polariton_2015}. They are a versatile building block for the realisation of enhanced sensors \cite{boriskina_spectrally_2006}, single photon sources\cite{liew_single_2010}, effective hamiltonian engineering \cite{sala_spin-orbit_2015} and may allow advances in photonic quantum simulations \cite{ritter_elementary_2012}, the study of Tonks-Girardeau gases \cite{carusotto_fermionized_2009} and many-body physics in general\cite{hartmann_strongly_2006}.

Much of the focus for microcavity research has been on monolithic solid state structures, such as micropillars grown by epitaxial methods. Recently however, an interest has emerged in open-access microcavities because of their facile tunability and flexibility to couple a range of electronic systems to the photonic mode \cite{barbour_tunable_2011, albrecht_narrow-band_2014, di_controlling_2012, miguel-sanchez_cavity_2013, trichet_open_access_2014, mader_scanning_2015, coles_diffusion-driven_2015, patel_gain_2015}. Open cavities bring the additional advantage over micropillar structures that  confinement of the optical mode can be achieved on sub-micron scales without either disrupting the integrity of the electronic system or causing scattering losses that inevitably occur with conventional etching and patterning techniques.
Adding to that, a high degree of control over the topography \cite{trichet_Topo_2014}, low mode volumes down to $\lambda^3$ and high finesses up to $2\times10^5$ were achieved with the current fabrication techniques \cite{hunger_fiber_2010,
dolan_femtoliter_2010,
uphoff_frequency_2015}. Recent demonstrations of polariton formation with quantum wells \cite{dufferwiel_strong_2014} and Purcell enhancements with 2D materials \cite{schwarz_two-dimensional_2014} have evidenced the maturation of the open-access microcavity design.
\newline \indent
In this work, we describe the first study of the engineered coupling between two of such open cavities forming a photonic molecule \cite{bayer_optical_1998,
boriskina_photonic_2010,
vasconcellos_spatial_2011}. We present the experimental formation of strongly coupled `supermodes' in the open-access geometry and model the results using a modemixing formalism. Two identical cavities are gradually merged moving from two uncoupled oscillators (features fully separated) to a single one (features fully merged). In between, the hybridisation of the fundamental mode as well as the first four transverse excited states is analysed. Moreover, taking advantage of the intrinsic tunability, we demonstrate the anti-crossing behaviour of this strongly coupled system by simply changing the angle between the two mirrors. This study paves the way to explore effects such as the unconventional polariton blockade \cite{verger_polariton_2006, bamba_origin_2011, xu_strong_2014} and create advanced structures such as rings and coupled arrays for future developments in the field of open-access nanophotonics.

\section{Chip design and experimental setup}
\label{setup}
\begin{figure}
\hspace{-10pt}
\includegraphics[width=0.5\textwidth]{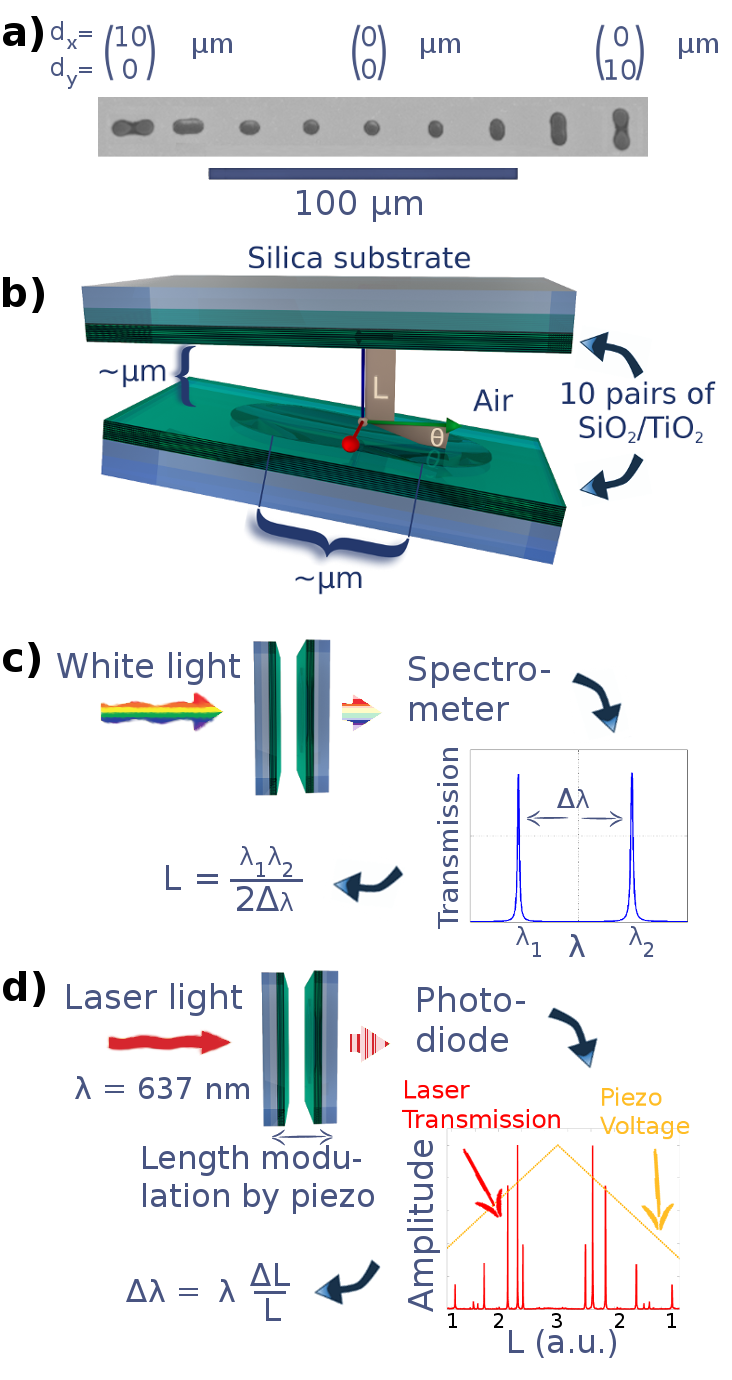}
\caption[flushleft]{\label{fig1} a) FIB micrograph of coupled cavities templates. From left to the right the distance $d$ between the two features is changed from $d = (10,0)~\SI{}{\micro\meter}$ over $(0,0)~\SI{}{\micro\meter}$ to $(0,10)~\SI{}{\micro\meter}$. \newline b) Experimental setup showing control over cavity length $L$ and angle $\theta$ between the two mirrors. c) Length initialisation routine with white light transmission. d) Laser transmission and equivalence between length and spectral sweep.\hfil\rule{\linewidth}{.4pt}}
\end{figure}

In the plano-concave configuration a tunable, open-access microcavity consists of two freely positionable, opposing mirrors, one flat and the other concave. Starting from flat Silica substrates from UQG Optics, we pattern one of them with a Focused Ion Beam (FIB) \cite{dolan_femtoliter_2010} in order to produce concave templates. Our fabrication method produces the desired shape with nanometric accuracy \cite{trichet_Topo_2014}. These templates are then coated with Distributed Bragg Reflectors (DBR's) at the Thin Film Facility of the Department of Physics (University of Oxford) via sputter deposition. In this work, we used 10 pairs of $\text{SiO}_2/\text{TiO}_2$ giving a macro-reflectivity above $99.7\%$ at the center wavelength of $\lambda =$~\SI{637}{\nano\meter}, corresponding to a maximum achievable finesse $\sim 1000$.

In order to study the gradual coupling behavior between two cavities, we produced chips with $13 \times 13$ features with varying inter-cavity distance $d$. The single feature shape was chosen to be the isophase surface of a Gaussian TEM$_{00}$ mode function (for further information on the exact shape and cavity merging parameters see Supplement 1). Fig. \ref{fig1} a) displays a FIB micrograph of a selection of these features on a silicon substrate, silica being non-conductive. In this example, the parameter $d$ is scanned from $d=$~\SI{10}{\micro\meter} along the $x$ axis (left cavity) to $d=$~\SI{10}{\micro\meter} along the $y$ axis (right cavity), going through fully merged cavity for $d=$~\SI{0}{\micro\meter} along both axes. The spectral properties of the fully merged cavities and separated ones are expected to be similar since they feature the same depth (\SI{400}{\nano\meter}) and radius of curvature ${R} =$~\SI{6}{\micro\meter}). Measurements were taken on a sample with $13 \times 13 = 169$ features, resulting in increments in the cavity separation parameter $d$ of $d_{step} =$~\SI{118}{\nano\meter}.

Once the two mirrors are brought close to each other within tens of microns, they are made parallel using a Thorlabs kinematic mount and the Fabry-P{\'e}rot fringes present in the planar part of the mirrors with an accuracy of \SI{200}{\micro\radian}.
The cavities are placed close to one of the rotation axes of the mount, since this configuration enables us to tune the relative angle between two adjacent coupled cavities without inducing significant cavity length changes (see Fig. \ref{fig1} b)). The setup is then initialised to a cavity length of $L \simeq$~\SI{3}{\micro\meter} by obtaining the free spectral range with white light transmission (see Fig. \ref{fig1} c)). The mode properties of individual cavities are now accessible with a fixed wavelength laser transmission experiment. By sweeping the cavity length with a piezo-microactuator the position of modes is measured relative to a defined ground state, thus enabling the conversion between cavity length and spectral position (see Fig. \ref{fig1} d)). The length scale for this sweep is obtained by traversing successive ground states, for which the cavity length differs by $\frac{\lambda}{2}$. After spatially selecting a single feature, each mode is identified by counting its nodes in $x$ and $y$ direction. Another way of obtaining this spectral and spatial information is used in the strong coupling demonstration, where we used the fluorescence of a dense layer of quantum dots to populate the cavity modes, which is conceptually similar to positioning a broad-linewidth, red lamp between the mirrors (for more detailed experimental description see Supplement 2).

\begin{figure*}[!ht]
\includegraphics[width=1.0\textwidth]{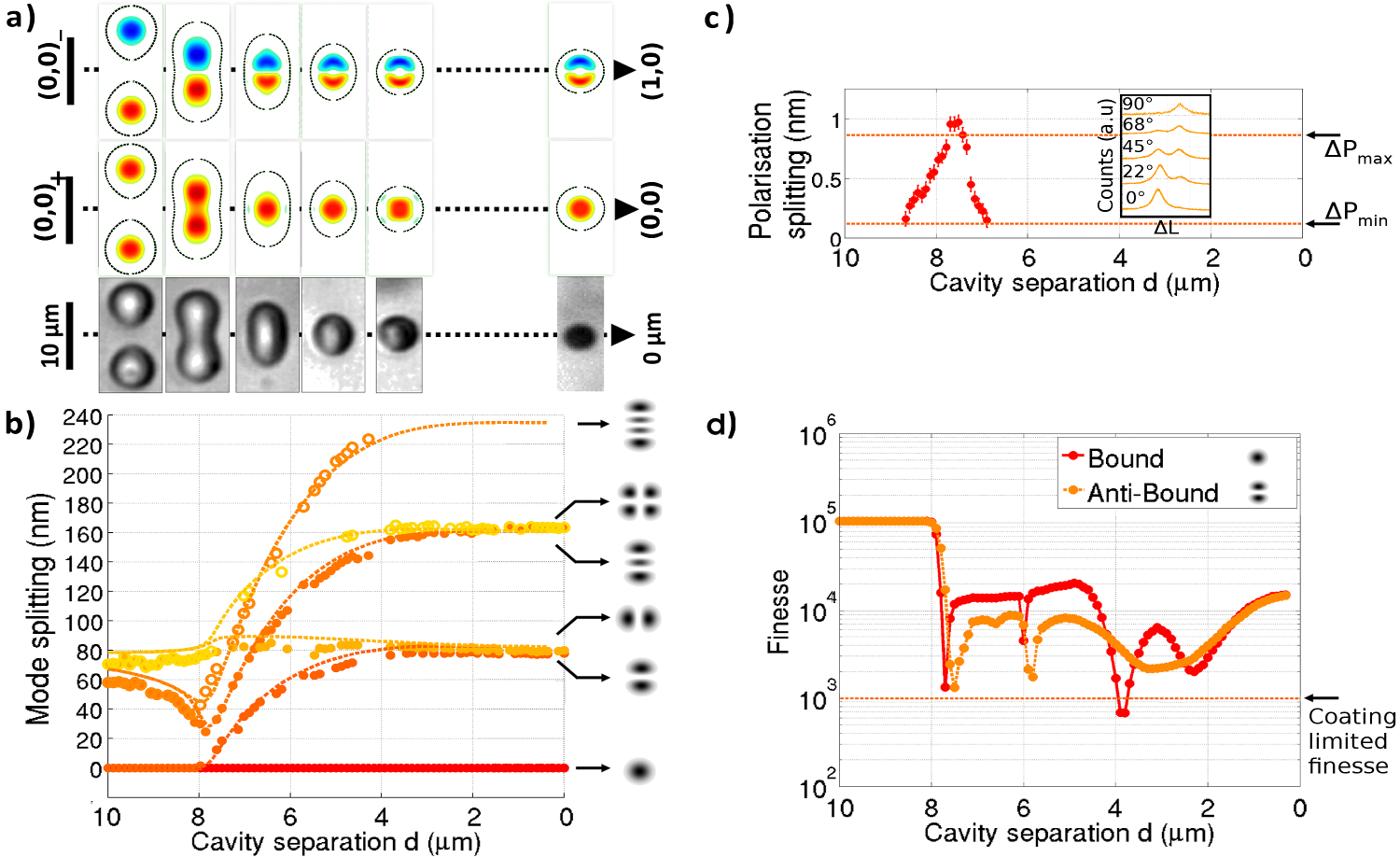}
\caption[foo]{\label{fig2} Mode hybridisation following the formation rule $((m,n)_+,(m,n)_-) $ $\rightarrow \left((2m,n), (2m+1,n) \right) $ for merging features. \newline
a) Real part of wavefunction showing the hybridisation of the first bonding and anti-bonding states (upper two rows), and microscope images of the respective coupled cavity feature (lower row). b) Experimental results (dots) obtained by laser transmission experiment and theoretical values (dashed lines) derived with modemixing formalism. Theoretical mode positions are shifted towards bigger $d$ values by \SI{350}{\nano\meter} (see text). c) Experimentally resolved polarisation splitting of first bonding state, leaving two peaks with mutually orthogonal polarisation (inset). $\Delta P_{min}$ denotes linewidth resolution limit, $\Delta P_{max}$ is calculated with formula (3) from \cite{uphoff_frequency_2015} for $\epsilon = 1$. d) Theoretical finesse values for first bonding and anti-bonding states obtained with modemixing formalism. The dips correspond to resonances with higher transverse modes of same $(m,n)$ parity for lower $q$ values (as reported in \cite{benedikter_transverse-mode_2015}).\hfil\rule{\linewidth}{.4pt}}
\end{figure*}

\section{Mode formation theory and modelling}
\label{theory}
The solutions of the paraxially approximated Helmholtz equation for cavities in the plano-concave geometry are the well-known Hermite-Gauss modes $\mathbb{H}_{q,m,n}(x,y,z)$. A state is defined by three natural numbers $q$, $m$ and $n$, where $q$ denotes the longitudinal and $m$, $n$ the transverse excitation quantum numbers, counting the number of nodes of the electric field in the respective planes. For an azimuthally invariant cavity those modes are $(m + n +1) $-fold degenerate but slight deviations lift the degeneracy and produce energy differences between the states. Each of these states has a further twofold polarisation degeneracy.

In the modemixing formalism introduced by Kleckner et al.\cite{kleckner_diffraction-limited_2010} solutions of the field distribution for a real cavity environment are found by solving the eigenvalue equation $\mathbb{M} \Ket{\Psi_i} =\gamma_i \Ket{\Psi_i}$, where $\mathbb{M}$ is the modemixing matrix which describes coupling between modes upon a single round trip within the cavity. The eigenvectors of $\mathbb{M}$ are the electric field amplitudes of stable cavity modes $\Ket{\Psi_i}$, where the eigenvalue $\gamma_i$ corresponds to the scaling factor the amplitudes undergo for each round trip in the cavity and $\Ket{\Psi_i}$ is a linear superposition of initial basis states $\Ket{\psi_j}$. A natural choice for $\Ket{\psi_j}$ is given by the normalised $\mathbb{H}_{q,m,n}(x,y,z)$ forming an orthonormal basis in the transverse plane \cite{kimel_relations_1993}. The spatial aspects of the cavity lead to the construction of $\mathbb{M}$, which is obtained by modulating the pairwise overlap of all basis states with the phase accumulated relative to a given plane due to the shape of the concave mirror $\Delta(x,y)$ (see Supplement 1). To introduce further aspects of the real cavity environment such as the angle between the mirrors, we deform $\Delta(x,y)$ by applying rotation matrices inducing a tilt by angle $\theta$ around the $x$ axis (Fig. \ref{fig1} b)).
It can be shown that the paraxial approximation holds for a single cavity with the parameters that define the basis states ($L \simeq$~\SI{3}{\micro\meter} and $R =$~\SI{6}{\micro\meter}). The same is true for the studied coupled cavity modes, since their waist is larger than in the single cavity case, thus guaranteeing a smaller divergence.

Now the full spectrum is obtained by looping over the cavity separation $d$ in small increments. At each $d$ value the spectrum of eigenvalues and eigenvectors of the corresponding modemixing matrix $\mathbb{M}$ is analysed, leading to the identification of the low energy modes in question. The spectral position (finesse) is obtained by analysing the phase (magnitude) of the eigenvalue $\gamma_i$ and the spatial wavefunction follows from the composition of the eigenvector $\Ket{\Psi_i}$ (Supplement 3 contains further information on this process).
  
\section{Results}
\subsection{Mode hybridisation}
\label{results1}
Figure \ref{fig2} shows the mode hybridisation process that links the cavity states while merging. The modes shown correspond to a longitudinal quantum number $q = 9$ or a cavity length around $L \simeq$~\SI{3}{\micro\meter}. The evolution of the modes from fully separated to merged is given by the rules $(m,n)_{+} \rightarrow (2m,n) $ and $(m,n)_{-} \rightarrow (2m+1,n)$. Here the subscripts $+$ and $-$ denote the in and out-of-phase relation between the two bare modes. The rule is illustrated in Fig. \ref{fig2} a), where the real part of the wavefunction for the first bonding and anti-bonding state is plotted alongside microscope images of the coupled cavity feature. Fig. \ref{fig2} b) depicts experimental and theoretical spectral information of the first six modes relative to the ground state $m + n = 0$. The premise that fully separated ($d =$~\SI{10}{\micro\meter}) and merged ($d =$~\SI{0}{\micro\meter}) cavities display the same mode structure is found to be only approximately correct for our system due to the long distance influence of the added shapes $z_\phi(x-d/2,y)$ and $z_\phi(x+d/2,y)$ on each other. For separations $d \rightarrow \infty$, the mode structure coincides. For $d=$~\SI{0}{\micro\meter}, the effective radius of curvature measured via the transverse mode splitting $\Delta^t L$ is given by $R = \frac{L}{\text{sin}^2 \left( \frac{2 \pi \Delta^t L}{\lambda} \right)}$ which gives a value of \SI{6}{\micro\meter}~$\pm$~\SI{0.1}{\micro\meter} in excellent agreement with the fabrication method input parameters. Starting with a cavity separation of $d =$ \SI{10}{\micro\meter} a pronounced splitting between the modes $(1,0)$ and $(0,1)$ is observed. The in- and out-of phase modes of the two branches stay degenerate down to $d =$~\SI{8.1}{\micro\meter}. For separation $d$ below \SI{8.1}{\micro\meter} the degeneracy is lifted and the coupling between bonding and anti-bonding state grows. For higher $n$ states this process sets in later and the splitting is less pronounced. In between $d =$~\SI{6}{\micro\meter} and $d =$~\SI{8}{\micro\meter}, the mode energy decreases for higher $m$ modes due to the decrease of the confining transverse potential.  For smaller $d$ the modes gradually evolve to the single deformed cavity case and display a gradually declining splitting with full recovery of the  $(m + n +1) $-fold degeneracy for $d =$~\SI{0}{\micro\meter}. 
Experimental and theoretical values show qualitative agreement with slight deviations in the horizontal position. This mismatch could be accounted for by shifting the theoretical curve in Fig. \ref{fig2} b) and d) by \SI{350}{\nano\meter} towards larger $d$ values. The same shift has been applied to all theoretical graphs throughout this paper. We propose that this mismatch stems from the neglected penetration depth into the DBR ($L_{\text{DBR}}\approx$~\SI{900}{\nano\meter}) in the modemixing theory. Theoretically a decrease of the effective length of the cavity leads to a decreased mode waist on the curved mirror and to smaller inter-modal coupling, effectively shifting the curve towards smaller $d$ values. With this compensation we obtain quantitatively well matched mode splittings across the whole $d$ parameter range. \\

In general each of these modes is twofold degenerate with respect to polarisation. Fig. 2 c) shows the lifting of this degeneracy for the first bonding state for intermediate $d$ values between $d =$ \SI{7.0}{\micro\meter} and $d =$ \SI{8.6}{\micro\meter}, where the eccentricity of the concave shape is maximal, leaving two well distinguishable cross-polarised peaks (see inset of Fig. \ref{fig2} c) ). Below and above these values no polarisation induced splitting is observed within the linewidth resolution limit given by the coating limited finesse values of $\approx$ 1000 (limit shown by $\Delta P _{min}$). The evolution of the splitting with $d$ tends to rule out a birefringence originating from the coating. In addition, we have checked over identical samples that the lower energy mode is always polarized perpendicularly to the coupling direction. We therefore attribute the splitting to nonparaxial field components along the longitudinal axis of the cavity as described in \cite{uphoff_frequency_2015}. This effect is large in our structure due to the small radius of curvature compared to reference \cite{uphoff_frequency_2015}. The maximal value of the splitting were taken from \cite{uphoff_frequency_2015}, and gives $\Delta P _{max}$ by evaluating $\Delta P = \frac{\epsilon^2 \lambda }{4 \pi k R}$ for the eccentricity $\epsilon = 1$. As seen in Fig. \ref{fig2} c), this prediction is in good agreement with our measurement for highly elliptical cavities ($d \simeq$~\SI{7.5}{\micro\meter}). At this position, the coupled cavity profile transitions from saddle shaped to elliptical. For both higher and lower $d$ values, the splitting rapidly decreases below the linewidth resolution limit.
\begin{figure}[!ht]
\includegraphics[width=0.47\textwidth]{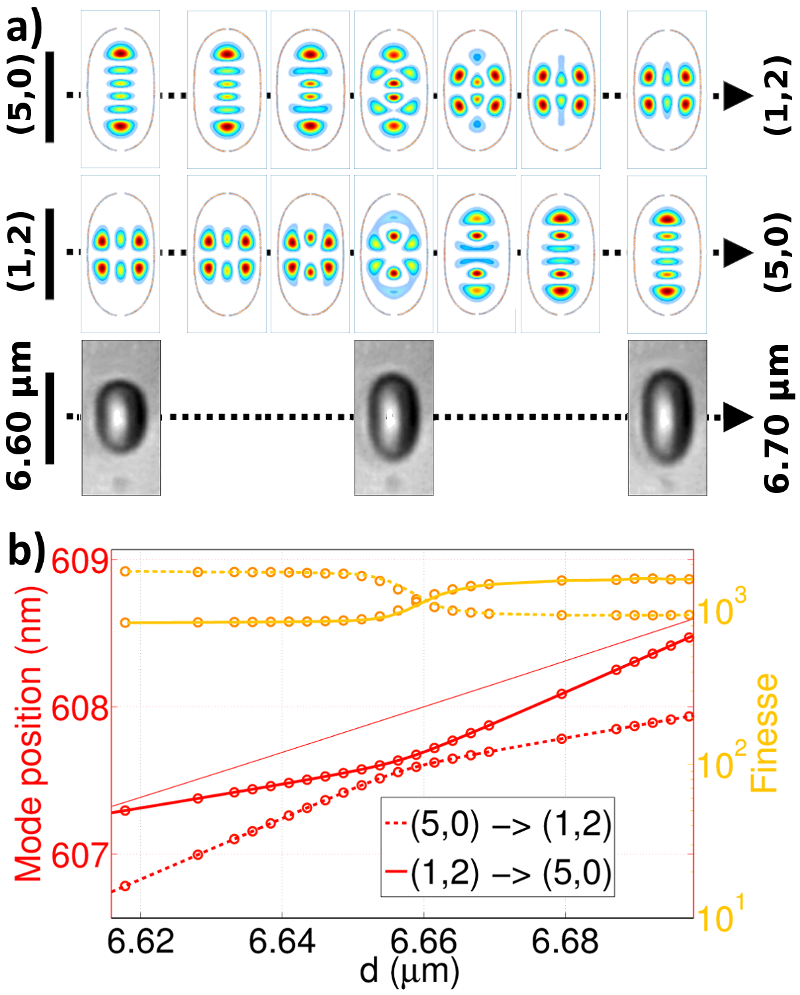}
\caption[foo]{\label{fig4} Anti-crossing of higher transverse cavity modes with same $(m,n)$ parity, $(1,2)$ and $(5,0)$, obtained by modemixing analysis as the cavity separation is tuned across the resonance position. a) Spatial wavefunction showing gradual interchange of the mode identity. b) Spectral position (left ordinate) and finesse (right ordinate) demonstrating the avoided level crossing. The third red line corresponds to mode $(3,1)$ in single cavity notation.\hfil\rule{\linewidth}{.4pt}}
\end{figure}

Fig. \ref{fig2} d) depicts finesse values of the first bonding and anti-bonding state. By application of the modemixing formalism each mode was assigned a finesse by equating $F_i \approx \frac{2 \pi}{1-\gamma_i^2}$, where $\gamma_i$ is the eigenvalue corresponding to the eigenmode in question \cite{kleckner_diffraction-limited_2010}. The experimental limit is given by the dashed line and shows that the coating is the limiting factor for our system. The higher valued diffraction limits show a decrease for intermediary $d$ values with several sharp dips in between before reaching maximal values for $d $ above \SI{7.5}{\micro\meter}. To this end we capped the maximal finesse values at $F = 10^5$ corresponding to mirror losses of 30 ppm because of the finite numerical accuracy given by the modemixing approach. The general trend of the values shown is a result of the long distance influence of the two added concave shapes causing larger feature sizes on the high $d$ end thus reducing diffraction losses. This effect vanishes for larger $d$ values where the finesse returns to the values reported for $d =$ \SI{0}{\micro\meter}. 
The intermediary dips occur at positions where the respective mode is in resonance with a higher transverse mode of lower $q$ which has the same parity. In this manner the bonding state $(q,0,0)$ has a dip at $d =$~\SI{3.85}{\micro\meter} where it is in resonance with the $(q-1,2,2)$ state. This phenomenon has recently been reported for single cavities by Benedikter et al. \cite{benedikter_transverse-mode_2015}. Experimentally the resolution of these dips was not possible due to the relatively low coating-limited finesse (dashed line in Fig. \ref{fig2} d)).

Following the mode evolution for higher excited transverse states the modemixing formalism predicts avoided crossings between modes of same $(m,n)$ parity. One of these anticrossings occurs between state $(1,2)$ and $(5,0)$ in single cavity notation around $d \simeq$~\SI{6.31}{\micro\meter} with a splitting of $\hbar \Omega =$~\SI{0.26}{\milli\eV}. Fig. \ref{fig4} shows the analysis of this phenomenon, depicting the spatial wavefunctions (a), the mode positions (b, left ordinate) and the mode finesses (b, right ordinate) as obtained with the modemixing formalism by looping over $d$ in \SI{2.5}{\nano\meter} increments. The gradual exchange of the mode identity given by its finesse and spatial wavefunction is accompanied by an avoided level crossing in the spectral domain.

\subsection{In-situ control of the coupling strength}
\label{results2}
\begin{figure}
\includegraphics[width=0.45\textwidth]{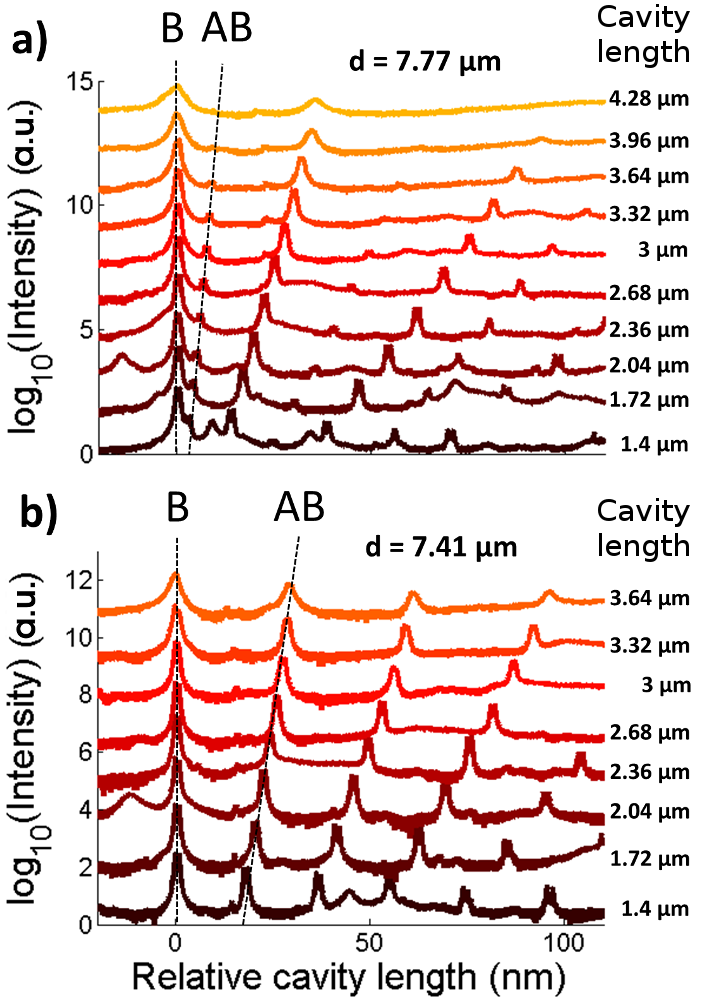}
\caption[foo]{\label{fig5} Logarithmic transmission spectra for different cavity lengths (equivalent to longitudinal mode number $q$) for a feature separation of a) $d =\SI{7.77}{\micro\meter}$ and b) $d =\SI{7.41}{\micro\meter}$. The experimental curves are shifted with respect to each other for clarity. The bonding (B) and anti-bonding (AB) states are highlighted with the black dashed lines. The absolute cavity length at the B state is given on the right hand side of the figure.}\hfil\rule{\linewidth}{.4pt}
\end{figure}

The wide length tunability of open-access microcavities allows a degree of in-situ control over the mode overlap and therefore inter-cavity coupling strength. In this set of experiments, the cavity length is scanned over the full stability range to extract spectra for coupled modes as a function of longitudinal mode number. Fig. \ref{fig5} a) and b) present these logarithmic spectra stacked on top of each other for $d =\SI{7.77}{\micro\meter}$ and $d =\SI{7.41}{\micro\meter}$. The bonding state mode position is used as a reference point set to \SI{0}{\nano\meter}. Each spectrum is labelled with its respective cavity length for this ground state on the right side of Fig. \ref{fig5} a) and b). The smallest optical cavity length achieved is $\SI{1.4}{\micro\meter}$ which is the smallest length attainable without putting the mirrors into contact (including the DBR penetration depth). Above $L \approx \SI{4}{\micro\meter}$, the cavity mode linewidth increases significantly because of diffraction losses \cite{kleckner_diffraction-limited_2010}. For $L = $\SI{3}{\micro\meter} the mode spacing corresponds to the values shown in Fig. \ref{fig2} b) for the respective $d$ parameter. In both Fig. \ref{fig5} a) and b), the bonding (B) and anti-bonding (AB) states are highlighted with black dashed lines. The splitting between these two states is increasing with the absolute cavity length. This is consistent with the fact that the mode waists, and therefore the mode overlap, is increasing for larger length leading to an increased inter-modal coupling. A similar behaviour can be observed for higher excited states as well. For the cavity with $d =\SI{7.77}{\micro\meter}$ (respectively $d =\SI{7.41}{\micro\meter}$), the splitting is then tuned from \SI{2.7}{\nano\meter} to \SI{8.6}{\nano\meter} (respectively \SI{17.7}{\nano\meter} to \SI{28.3}{\nano\meter}) by steps of 0.8 nm (respectively 1.7 nm).

\subsection{Strong coupling}
\label{results3}
\begin{figure}
\includegraphics[width=0.5\textwidth]{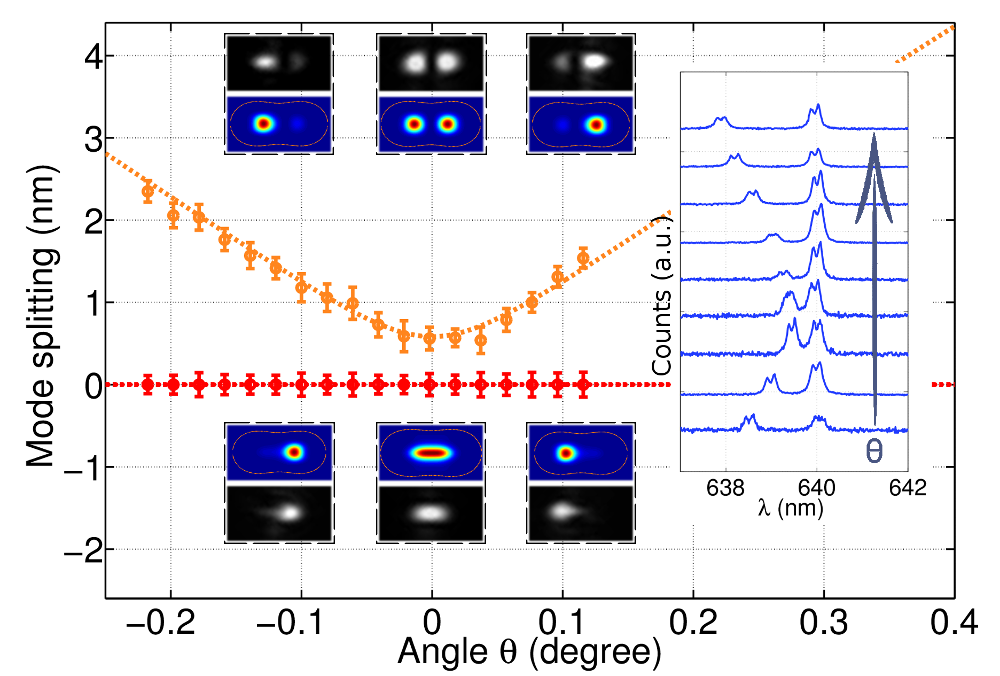}
\caption{\label{fig3} Mode splitting of ground $(0,0)$ and first excited $(1,0)$ (bonding and anti-bonding) state for coupled cavity with $d =$ \SI{7.45}{\micro\meter} showing anticrossing behaviour when tuning the angle between the two mirrors (see \ref{fig1} b)). The dots correspond to experimental results with linewidths given by the error-bars and agree well with the theory (lines). The small insets depict the spatial wavefunction for the two modes as observed experimentally (grey colourscale) and predicted theoretically via modemixing (blue-red colourscale). The large inset shows a selection of the original spectra;  each mode is polarisation split due to the eccentricity of the cavity (comp. with Fig. \ref{fig2} c)).  }\hfil\rule{\linewidth}{.4pt}
\end{figure}

In order to prove that the cavities are strongly coupled and that the spectral splittings are not just an effect of the cavity shape, we demonstrate the characteristic anti-crossing behaviour between the ground state and first excited state. This experiment is performed simply by changing the angle between the two mirrors, thus varying the relative length of the two cavities. This experiment was performed for a separation parameter $d =$~\SI{7.45}{\micro\meter} and for the angular range $-0.25^{\circ} < \theta < 0.15^{\circ} $ (see Fig. \ref{fig1} b) for visualisation). For each angle, the length of the cavity was adjusted to fix the ground state $n+m=0$ at the spectral position $\lambda = $~\SI{640}{\nano\meter}. Each mode was split by \SI{0.3}{\nano\meter} into two orthogonally polarised peaks due to the cavity eccentricity and was fitted with a double Lorentzian function (see the large inset in Fig.~\ref{fig3}). The width and the position of the two peaks showed the same behaviour and contain the same information, which is plotted in Fig. \ref{fig3} as a function of the angle $\theta$. The dots denote the mean value of the fitted peaks and the error-bars give their respective linewidths. The continuous line is the result obtained with the modemixing formalism with rotation matrices $\mathbb{R}_x(\theta)$ applied to modulate the mirror shape $\Delta(x,y)$. The anticrossing behaviour is clearly observed. For Fig.~\ref{fig3}, the strength of the coupling, defined by the ratio of the mode splitting $\Omega$ and the linewidth of the modes $\Gamma$, gives a value of $f = \frac{\Omega}{\Gamma} = 2$, demonstrating the strong coupling character of the system. For large separation $d$, the strong coupling regime vanishes for $f = 1$ which corresponds, in our case to a separation of $d =$~\SI{8.1}{\micro\meter}. Further pictorial confirmation of the accurate description by the modemixing formalism is given by the spatial wavefunctions obtained theoretically (red-blue colourscale insets) and experimentally (grey colourscale insets). 
The images show clearly the hybridisation between individual cavity modes for tuning through the avoided crossing. The same behaviour is seen for higher excited supermodes, which show less pronounced angular sensitivity the higher the excitation number $m$ in the elongated direction is.

\section{Conclusions and Outlook}

In this work we have demonstrated control over coupled open-access microcavities with mode volumes around $10~\times~\lambda^3$. We presented experimental results from coupled cavities in which the inter-cavity spacing is varied and compared the results with predictions obtained by a modemixing formalism. Furthermore, the strongly coupled character of the bonding and anti-bonding states is demonstrated by the mode anti-crossing. By making use of the degrees of freedom which the open-access cavity provides in-situ (length and angular tuning), the coupling between individual cavity modes of the system can be controlled. 

This work shows the ability to engineer the mode structure of coupled open-access cavities as a fundamental building block for photonic devices. Such photonic molecules have direct implications for the study of coupled quantum systems. For example they are attractive for the study of polariton blockade effects and the creation of ultrafast single photon sources. In addition the control of the photonic potential offers the prospect for engineering the Hamiltonian of coupled Bose-Einstein-Condensates \cite{nyman_interactions_2014}. Such studies could be applied to both exciton-polariton and photonic systems.

\begin{acknowledgement}
We thank Richard Makin at the Thin Film Facility (TFF) of the Department of Physics (University of Oxford), Gareth Hughes for technical support with the FIB and Paul Warren for computer cluster access. LF and AT acknowledge funding from the Leverhulme Trust. The authors declare no competing financial interest. See supplementary material for supporting content.
\end{acknowledgement}

%
\bibliography{ccpaper}

\newpage
\begin{center}
\textbf{\large Supplemental Materials}
\end{center}

\setcounter{section}{0}
\setcounter{equation}{0}
\setcounter{figure}{0}
\setcounter{table}{0}
\makeatletter
\renewcommand{\thesection}{S\arabic{section}}
\renewcommand{\theequation}{S\arabic{equation}}
\renewcommand{\thefigure}{S\arabic{figure}}

The shape of the concave feature forming one side of the coupled cavity is given analytically. Additionally the details of the experimental configuration and devices used to record the presented data in the main text is discussed. Finally the technical details of the modemixing formalism and the entailed computational methods are exposed.

\section{Chip design}
\label{chip_design}

We chose the shape of a single cavity to be the isophase surface of a Gaussian mode \cite{kimel_relations_1993}. We keep this formulation for radial displacements that extend well beyond the paraxial approximation. This approach provides us an analytical form as well as a smooth shape which is ideal to prevent any crack formation in the DBRs. The phase of a Gaussian cavity mode at a position $(x,y,z)$ is given by:
\begin{align}
\phi (x,y,z) = \frac{k(x^2+y^2)}{2 R(z) } -\text{atan}\left(\frac{z}{z_R}\right) + k z 
\label{isophase}
\end{align}
where $R(z)$ is the radius of curvature (RoC) for $x=y=0$ and is given by $R(z)=z \left[1+\left(\frac{z_R}{z}\right)^2 \right]$ , $z_R=(\pi w_0^2)/\lambda$ is the Rayleigh length and $k$ is the mode wavevector  given by $k=2\pi/ \lambda $. In order to confine the light, the cavity length $L$ has to be smaller than the radius of curvature $R$ on axis. In this case, the waist of a Gaussian mode is given by $w_0^2=~\frac{\lambda L}{\pi} \sqrt{\frac{R}{L}-1}$. In this work, we have used a RoC of \SI{6}{\micro\meter} for a targeted cavity length in the middle of the stability range, ie. $L =$~\SI{3}{\micro\meter}. With this set of parameters, the waist is of the order of \SI{800}{\nano\meter}, which leads to a single cavity mode volume of $44 \times \left( \frac{\lambda}{2} \right)^3$. To obtain the surface coordinate $z_\phi$ as a function of $(x,y)$ of the isophase surface, one has to solve the following equation:
\begin{align}
\label{eq:phase2solve}
\phi(x,y,z_\phi)=\phi(0,0,L)=kL-\text{atan} \left( \frac{1}{\sqrt{\frac{R}{L}-1}}\right)
\end{align}
The coupled cavity shape $\Delta(x,y)$ is constructed by adding two surfaces defined by the function $z_\phi(x,y)$ separated by a distance $d$ (blue dashed lines on Fig. \ref{fig1} a)) which leads to the equations:
\begin{align}
\Delta(x,y) = Z(x,y) \frac{L}{Z^{max}}
\label{ccshape_1}
\end{align}
\begin{align}
Z(x,y) = z_\phi(x-d/2,y) + z_\phi(x+d/2,y)
\label{ccshape_2}
\end{align}

where $\frac{L}{Z^{max}}$ is a normalization term that keeps the maximum length of the coupled cavities constant. As long as there is a saddle point in between the two cavities, $Z^{max}\approx L$ and the normalization factor is close to unity. However, when the saddle point disappears, the normalization factor keeps a constant length for the new structure created. For the cavities presented in this article, these two cases are delimited by a separation $d=$~\SI{7.6}{\micro\meter}. We use the shape given by the function $\Delta(x,y)$ as an input of our fabrication method. The lateral extension of the feature is dictated by its depth that we choose to be 400 nm. For a single cavity, a depth of \SI{400}{\nano\meter} corresponds to a diameter around \SI{5.5}{\micro\meter}, which is 7 times larger than the expected waist. Therefore, above $99.99\%$ of the mode is contained within the feature. The normalization term in equation \ref{ccshape_1} enables the comparison between each feature since the respective fundamental mode energies will be close to each other.

\begin{figure}
\hspace{-10pt}
\includegraphics[width=0.48\textwidth]{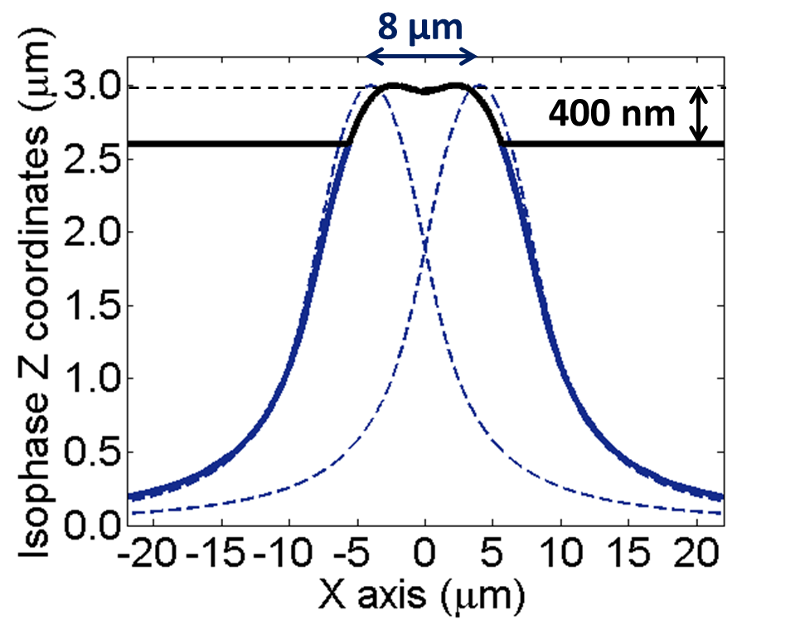}
\caption[flushleft]{\label{fig1} Gaussian mode isophase. The dashed lines correspond to the isophase for two independent cavities separated by \SI{8}{\micro\meter}, the solid blue line to the sum of each independent isophase (eq. \ref{ccshape_2}) taking the normalisation factor into account. The black solid line gives the patterned shape which includes only the last \SI{400}{\nano\meter} of the coupled cavity shape.\hfil\rule{\linewidth}{.4pt}}
\end{figure}

\section{Experimental setup}
\label{exp_setup}

Initially the absolute cavity length is set to \SI{3}{\micro\meter} using a white light transmission experiment (Bentham WLS100). The collected light is then focused on an Andor combined spectrograph/CCD with a 300 grooves/mm grating in order to record the transmission spectrum.

We use laser transmission to study the spectral and spatial mode structure of the cavities. A continuous wave TOPTICA laser DL100 with a fixed wavelength of $\lambda =$~\SI{637}{\nano\meter} (linewidth $< 1~$MHz) is focused with an aspheric lens with ($\text{focal length} =$~\SI{5}{\centi\meter}) onto the chip. In this way the spot size is always bigger than the cavity contour allowing mode matching even to the higher transverse modes discussed in the main text. The chip is then imaged onto a Andor Newton CCD or an avalanche photodiode from Thorlabs (ref: APD120A2/M). One of the mirrors is mounted onto a ring piezo-actuator from Piezomechanik in order to sweep the cavity length. The length scale for the cavity sweep is obtained by traversing through successive ground states with a well known length separation of $\frac{\lambda}{2}$. By spatially selecting specific cavities, we can record the transmission spectrum as a function of cavity length. With this method, each maximum transmission peak is associated with a spatial wavefunction imaged onto a CCD allowing us to count the number of nodes and anti-nodes along the x and y direction, and therefore, properly identify each mode. Since the fundamental mode energy changes significantly as a function of d, we do not have an absolute measurement of the cavity length. Therefore, each mode observed is referred to with its energy difference from the ground (bound) state.

A different method of obtaining single cavity spectra is used in the strong coupling demonstration, where the angle between the two mirrors is varied. Here we embed an ensemble of core-shell CdSe/ZnS quantum dots (Lumidot 640 from Sigma Aldrich) into a thin spin-coated PMMA film on the planar mirror. After initialising the setup to a similar cavity length as above ($L \simeq$~\SI{3}{\micro\meter}) with a white light transmission experiment, a PicoQuant LDH-D-C-470 laser with $\lambda =$~\SI{470}{\nano\meter} and continuous wave mode is focused at a single cavity with moderate power densities of $\simeq$~1000~W/cm$^2$ to excite the quantum dots and cause emission into the cavity modes around $\lambda =$~\SI{640}{\nano\meter}. By adapting the cavity length such that the $m+n = 0$ mode is positioned at $\lambda =$~\SI{640}{\nano\meter} the relative mode spacing to higher excited states is determined with lorentzian fits to the mode peaks in the cavity emission spectrum, which is recorded with a 1800 grooves/mm grating. 

\section{Modemixing formalism}

In the modemixing formalism stable cavity modes are found by solving the eigenvalue problem $\mathbb{M} \Psi_i = \gamma_i \Psi_i$, where $\mathbb{M}  \in \text{Mat}(n,n)$,  $i \in {1,2,...,n}$ and $\Psi_i$ is an eigenvector with its corresponding eigenvalue $\gamma_i$. 
The modemixing matrix $\mathbb{M}$ is constructed according to \cite{kleckner_diffraction-limited_2010} eq. (4) to (6). Each eigenmode $\Ket{\Psi_i}$ is a linear superposition of initial basis states $\Ket{\psi_j}$. A natural choice for the $\Ket{\psi_j}$ is given by the normalised Hermite-Gauss basis states $\mathbb{H}_{q,m,n}(x,y,z)$ as defined in \cite{kimel_relations_1993}. Here $j \in {1,2,...,n}$ maps each dimension of the original modemixing matrix to a combination of transverse excitation numbers $m,n$.
For our purposes a set of the first 500 basis states $\mathbb{H}_{q,m,n}(x,y,z)$ ordered by increasing $m+n$ value is chosen. Due to the elongation of the cavity profile in one dimension we truncate the basis set at $n+m = 15$ in the perpendicular direction allowing modes of order up to $n+m = 38$ in the elongated direction. For feature separations above $d =$ \SI{8}{\micro\meter} we change the origin of the basis set to co-align with the left feature. Care was taken that both methods produced the same results for an overlapping region \SI{8}{\micro\meter} $ < d < $ \SI{8.5}{\micro\meter}. The convergence of the process is evident by the fast decaying contributions of higher order modes in the composition of the eigenstates of $\mathbb{M}$, leading to an accurate representation of the actual field distribution. The exact calculation of the matrix elements is facilitated by the symmetry of the system, allowing the integration to be done in one quadrant only. The method is based on MATLABs quad2d routine with absolute and relative error bounds of $10^{-12}$ and $10^{-10}$ respectively running on a parallel pool with $4 \times 8$ workers each running at 2.6~GHz. \\
We infer the mode wavelength from the phase $\Delta \varphi_i $ of the eigenvalue by noticing that $ \lambda_i = \frac{\lambda}{1 + \frac{\Delta \varphi_i \lambda}{4 \pi L}}$, where $\lambda$ and $L$ are the wavelength and cavity length defining the original basis states. The mode finesse is obtained by equating $F_i \approx \frac{2 \pi}{1-\gamma_i^2}$, which corresponds to diffraction losses only  \cite{kleckner_diffraction-limited_2010}.
This procedure leads to spectral and spatial determination of the mode properties. The full spectrum of the different cavity separation states is obtained by looping over $d$ in \SI{250}{\nano\meter} increments with a finer mesh of \SI{62.5}{\nano\meter} in the coupled regime for \SI{6.8}{\micro\meter} $ < d < $ \SI{8}{\micro\meter}.

\end{document}